\documentclass[11pt,twoside]{article}
\usepackage{./asp2014}
\usepackage{subfig}
\usepackage{graphicx}
\usepackage[table,x11names]{xcolor}

\aspSuppressVolSlug
\resetcounters

\begin{document}

\title{On the Incidence Rate of Blazhko Stars}
\author{Geza~Kovacs  
\affil{Konkoly Observatory, Budapest, 1121 Konkoly Thege ut. 13-15, Hungary \email{kovacs@konkoly.hu}}
}

\paperauthor{Sample~Author1}{Author1Email@email.edu}{ORCID_Or_Blank}{Author1 Institution}{Author1 Department}{City}{State/Province}{Postal Code}{Country}

\begin{abstract}
In a recent paper \citep{2018A&A...614L...4K} we examined the 
incidence rate of the modulated RR Lyrae stars by using the data 
from Campaigns 01$-$04 of the Kepler K2 mission. We found an 
observed rate of $\sim 90$\%, implying near $100$\% underlying rate, 
after correcting for detection bias due to observational noise. In 
this work we extend the sample to Campaign 08 and check the reliability 
of our earlier estimate. We get the same high rate, stressing the 
importance of full time series modeling (including systematics) in 
searching for shallow signals in the presence of large amplitude 
variabilities. 
\end{abstract}

%
%
\section{Introduction}
It is not easy to find an astrophysical phenomenon with as little 
clue to the physical nature of the cause of the phenomenon as  
the (quasi)periodic modulation of the light curves RR~Lyrae stars 
(or, as commonly known, the Blazhko effect -- \citealt{1907AN....175..325B}).
Although several ideas have been suggested, none of them succeeded, 
likely because of the missing underlying physics in those ideas 
-- see Koll\'ath in these proceedings and also earlier reviews of 
\cite{2016CoKon.105...61K} and \cite{2016pas..conf...22S}. Under this 
circumstance, most of the works in this field focus on further 
analyses of observational data to spark some workable idea. In 
this purely experimental approach various studies aim for finding 
new frequency components, establishing relations among the observed 
parameters (e.g., \citealt{2020MNRAS.494.1237S}) and derive 
accurate population parameters, such as incidence rates. In this 
work we follow the latter thread by extending our earlier work on 
the incidence rate of Blazhko stars \citep{2018A&A...614L...4K}.

%
%
\section{Method of Analysis}
The hunt for transiting extrasolar planets has revolutionized the 
analysis of photometric time series. Considering the small signals 
to be searched for, filtering out systematic effects both from 
the ground- and space-based observations has become of vital 
importance. Many methods have been developed primarily for transiting 
planet search, e.g., TFA, SysRem, PDC, SFF, EVEREST, respectively, 
by \cite{2005MNRAS.356..557K}, \cite{2005MNRAS.356.1466T}, 
\cite{2012PASP..124..985S}, \cite{2014PASP..126..948V}, 
\cite{2018AJ....156...99L}. However, only a few of these were 
extended to general variability search \citep{2016MNRAS.459.2408A}, 
primarily because of the difficulties encountered in separating the 
instrumental and 
environmental systematics from the underlying astrophysical signal. 
In particular, except for the very recent study by 
\cite{2019ApJS..244...32P}, all previous investigations on 
the Blazhko phenomenon have been carried out on datasets using 
Simple Aperture Photometry (SAP) for the space data 
\citep{2014ApJS..213...31B} and standard ensemble 
photometry for the ground-based data (e.g., analyses performed on 
the OGLE database -- see \citealt{2017MNRAS.466.2602P}).     

The main reason why standard methods (e.g., SFF by 
\citealt{2014PASP..126..948V}) are not applicable to variable stars 
with dominating large amplitude variations is that these methods 
employ the ``null-signal'' assumption.\footnote{That is, there is 
no underlying variation, except for the brief moments of the transit 
events.} Although inclusion of Gaussian process models has been 
proven to be a useful way of conserving stellar variability 
\citep{2016MNRAS.459.2408A}, here we resort to another approach, 
we have already employed in our pilot survey on Blazhko stars in 
fields C01-C04 of the K2 mission \citep{2018A&A...614L...4K}.

%
\begin{figure}[!h]
 \vspace{-10pt}
  \begin{minipage}[c]{0.55\textwidth}   
    \includegraphics[angle=-0, width=1.0\textwidth]{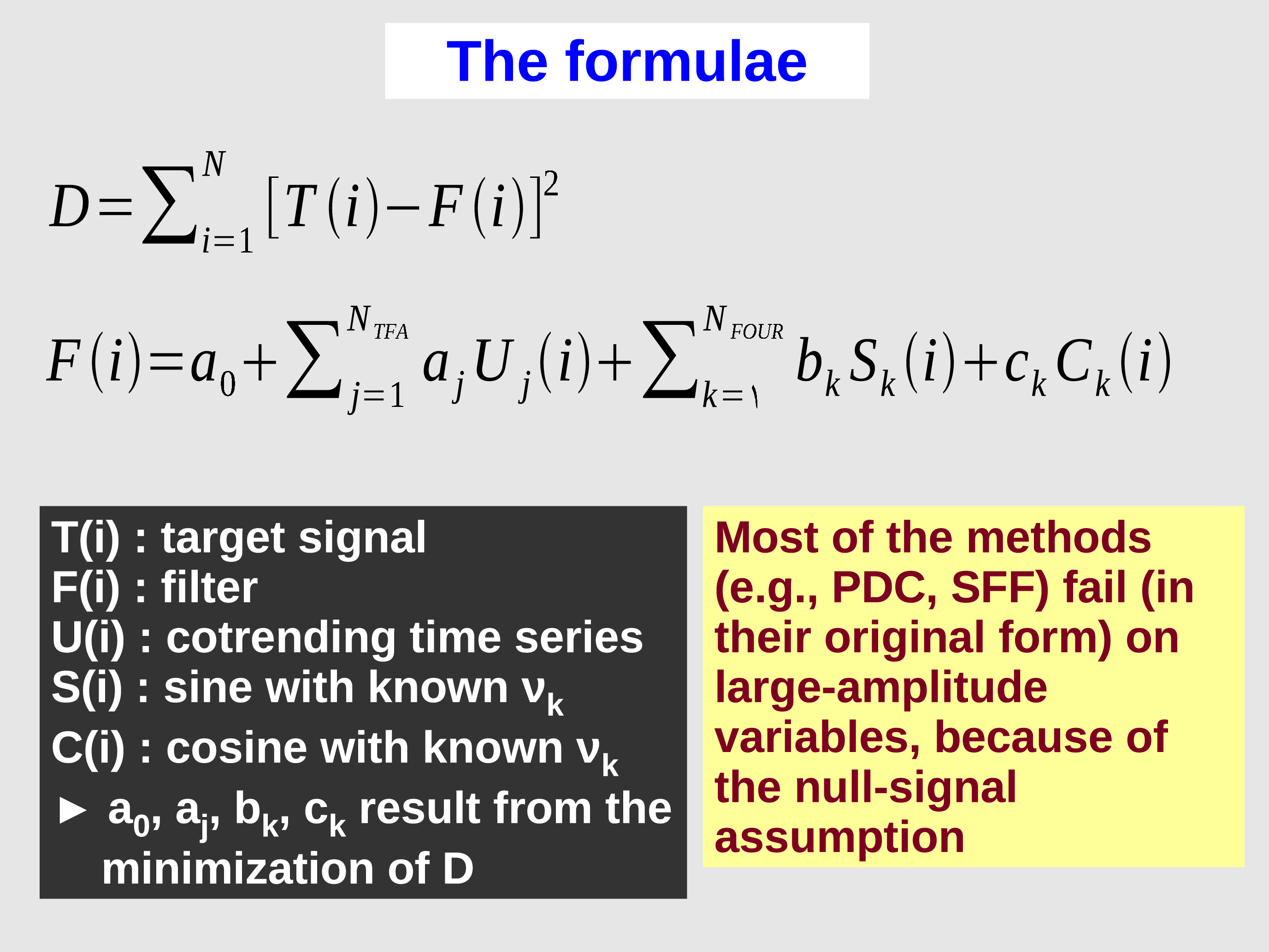}
  \end{minipage}\hfill
  \begin{minipage}[c]{0.40\textwidth}
\vspace{10pt}
\caption{Basic constituents of the time series model used to search for 
      small signal components. It is assumed that a good approximation 
      is available for the frequencies entering in the Fourier 
      representation of the large amplitude pulsation. We use a $15^{\rm th}$  
      order Fourier sum with $100$--$400$ cotrending time series 
      to compute systematics-free residuals to be searched for faint 
      signals.~Normalized fluxes are used throughout the data processing.}
    \label{equations}
  \end{minipage}
 \vspace{-10pt}
\end{figure}

As briefly summarized in Fig.~\ref{equations}, the observed signal 
is represented by the sum of two distinct components: i) the large 
amplitude pulsation, modelled by the Fourier series $\{b_k,c_k\}$ 
and ii) systematics, estimated by the linear combination of the 
time series $\{U_j(i)\}$. This latter set may include other 
photometric time series from the same field (or, the combinations 
thereof -- often referred to as cotrending vectors -- e.g., 
\citealt{2012PASP..124..985S}) and stellar image parameters (e.g., 
positions on the CCD chip). While using cotrending vectors enables 
us to grab the common features of the light variation (the very 
essence of systematics), the image parameters help to disentangle 
those variations in the stellar flux that have their origin in the 
peculiarities of the given stellar image \citep{2010ApJ...710.1724B}. 
For cotrending time series we select bright stars uniformly distributed 
throughout the field of the campaign under scrutiny. Whenever 
available, the cotrending set is extended by the pixel coordinates 
of the image center. 

%
%
\begin{figure}[!h]
 \vspace{15pt}
  \begin{minipage}[c]{0.55\textwidth}   
    \includegraphics[angle=-0, width=1.0\textwidth]{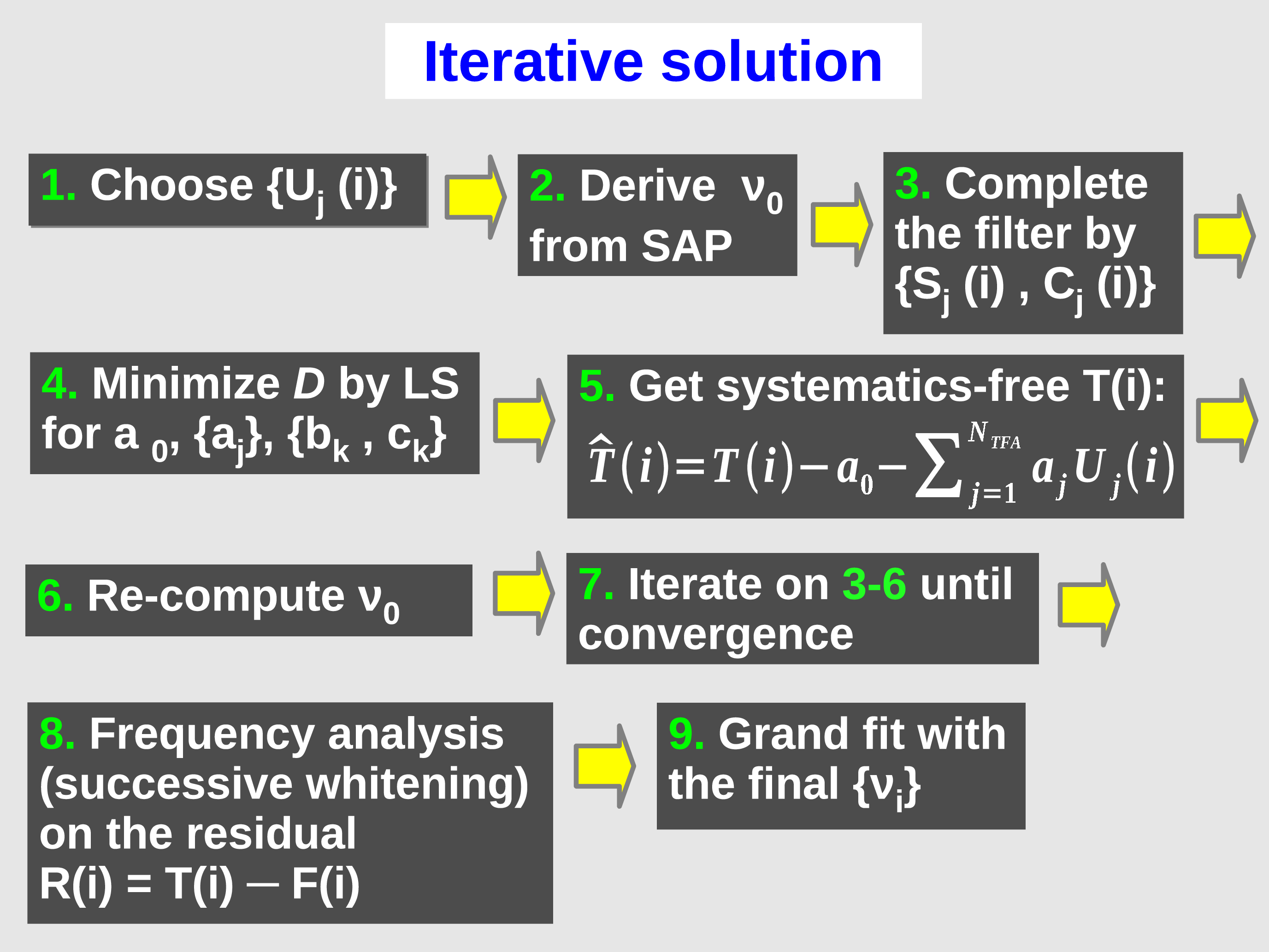}
  \end{minipage}\hfill
  \begin{minipage}[c]{0.40\textwidth}
\vspace{-30pt}
\caption{Flow chart of the algorithm employed in this work to separate 
         the large amplitude component and the systematics from the 
	 observed light curve $\{T(i)\}$ and access the small amplitude 
	 residual $\{R(i)\}$ to search for additional Fourier components. 
	 See Fig.~\ref{equations} and text for additional discussion of 
	 the symbols and the method.}
    \label{flow_chart}
  \end{minipage}
\end{figure}

Because the pulsation frequency is known only approximately from the 
direct analysis of the SAP data, we need to perform an iterative search 
to make the frequency determination more accurate. This procedure serves 
to avoid artificial remnant power in the residuals, that could be 
incorrectly identified as a modulation component. The multistep procedure 
is depicted in Fig.~\ref{flow_chart}. In the final grand fit (step 9) 
we also include the newly found small amplitude components to minimize 
the effect of incomplete signal modeling. The main interest of this 
paper is step 8, the search for additional frequency components, and, 
in particular, the search for those components that are close to the 
pulsation frequency (and its harmonics).   

%
%
\section{Results} 
By following the RR~Lyrae list of \cite{2018A&A...620A.127M}, 
as in our former study, we gathered the publicly available items from 
the NASA Exoplanet Archive\footnote{\url{https://exoplanetarchive.ipac.caltech.edu/})}. 
We relied mostly on the database stored in the ExoFOP section of the 
site by \cite{2015ApJ...811..102P} (see also \citealt{2016MNRAS.459.2408A}). 
When the target was missing in the ExoFOP section, but it was 
available through the Kepler pipeline \citep{2012PASP..124.1000S}, 
we used this latter set. In both cases only the respective raw 
(i.e., SAP) fluxes were used with TFA for correcting systematics. 
As in our earlier work, the classification of the variables 
was based on the residual frequency spectra: a star was declared 
as a Blazhko variable, if there was one or more significant 
peaks in the residual spectrum near the fundamental frequency. 

%
%
\definecolor{magicmint}{rgb}{0.67, 0.94, 0.82}
\definecolor{almond}{rgb}{0.94, 0.87, 0.8}
\definecolor{blizzardblue}{rgb}{0.67, 0.9, 0.93}
\definecolor{paleaqua}{rgb}{0.74, 0.83, 0.9}
\begin{table}[!h]
\centering
  \caption{Result of the search for modulation components.}
  \label{incidence}
  \scalebox{1.0}{
  \begin{tabular}{crrrrc}
  \hline
   Field    & N$_{\rm tot}$ & N$_{\rm RRab}$  & $N_{\rm BL}$ & $N_{\rm nonBL}$ &  $N_{\rm BL}/N_{\rm RRab}$ \\ 
 \hline\hline
 01  &   14   &   14   &    9 &  5\hskip 5pt  &  0.64 \\
 02  &   57   &   57   &   52 &  5  &  0.91 \\
 03  &   62   &   62   &   59 &  3  &  0.95 \\
 04  &   56   &   33   &   30 &  3  &  0.91 \\
 05  &   57   &   57   &   52 &  5  &  0.91 \\
 06  &  142   &  107   &   99 &  8  &  0.93 \\
\rowcolor{paleaqua}   07  &  353   &  235   &  184 & 51  &  0.78 \\
 08  &   47   &   46   &   44 &  2  &  0.96 \\
\hline
Sum: &  788   &  611   &  529 & 82  &  0.87 \\
\hline
\end{tabular}}
\begin{flushleft}
{\bf Notes:}~{\small 
N$_{\rm tot}:$ Total number of RRab stars from the list of 
\cite{2018A&A...620A.127M}; 
N$_{\rm RRab}:$ Actual number of RRab stars publicly available from the 
NASA Exoplanet Archive; 
$N_{\rm BL}:$ Number of Blkazhko stars; 
$N_{\rm nonBL}:$ Number of non-Blazhko stars (single period or other). 
Campaign $07$ is highlighted because of the low Blazhko rate -- see text.}
\end{flushleft}
\end{table}

The campaign-by-campaign incidence rates are shown in Table~\ref{incidence}. 
We missed certain targets from the list of \cite{2018A&A...620A.127M}, 
because they were not included in the publicly available databases. 
Nevertheless, this study is based on four times larger sample than 
our pilot survey \citep{2018A&A...614L...4K}, yielding 
statistically more significant estimate of the high incidence 
rate of the Blazhko effect. The field of C07 has apparently a low rate, 
but this is likely due to the overabundance of faint, blended 
targets associated with the Sagittarius dwarf galaxy.  

%
%
\begin{figure}[!h]
    \centering
    \subfloat{\includegraphics[angle=-0, width=0.45\textwidth]{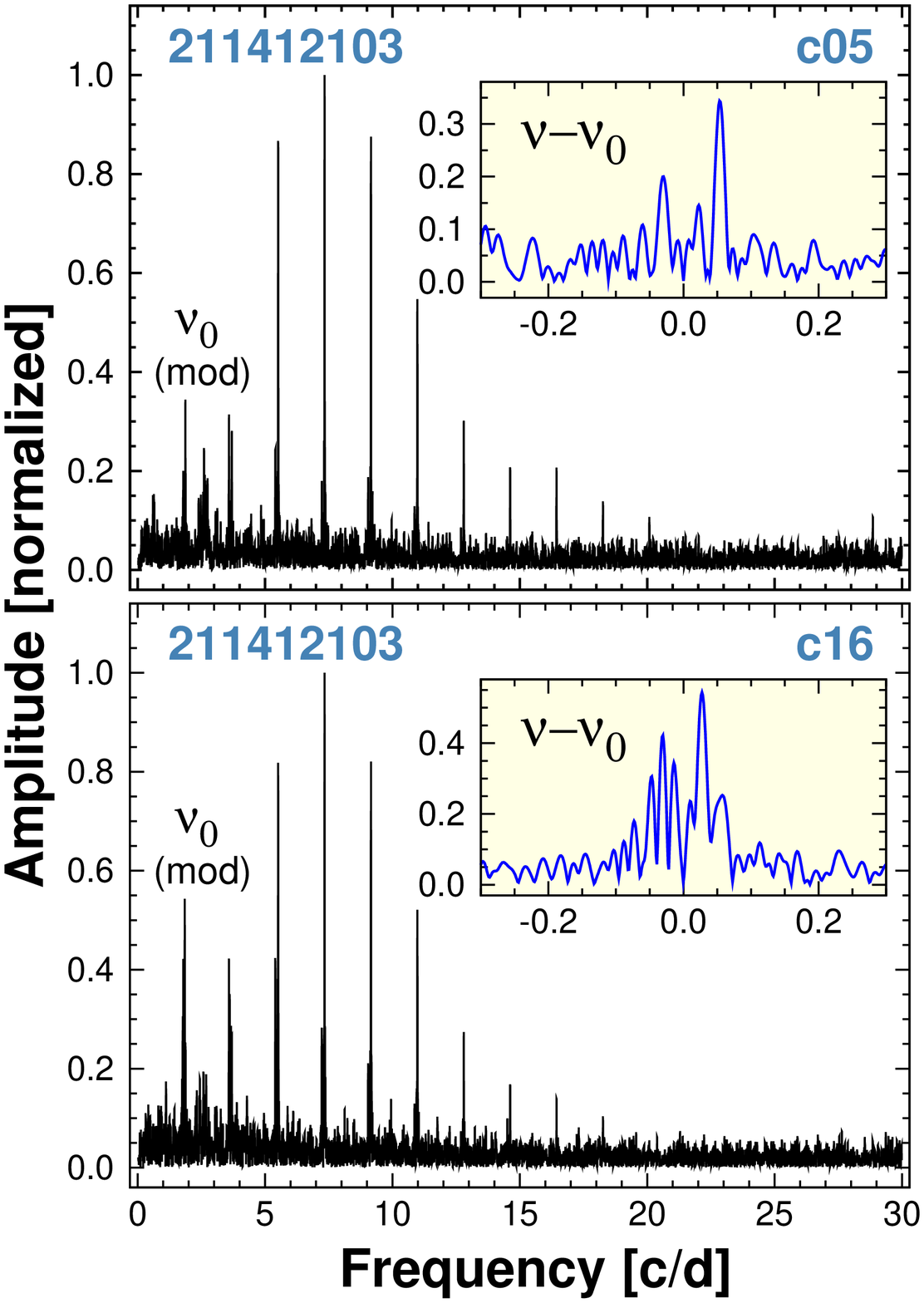}}
    \qquad
    \subfloat{\includegraphics[angle=-0, width=0.45\textwidth]{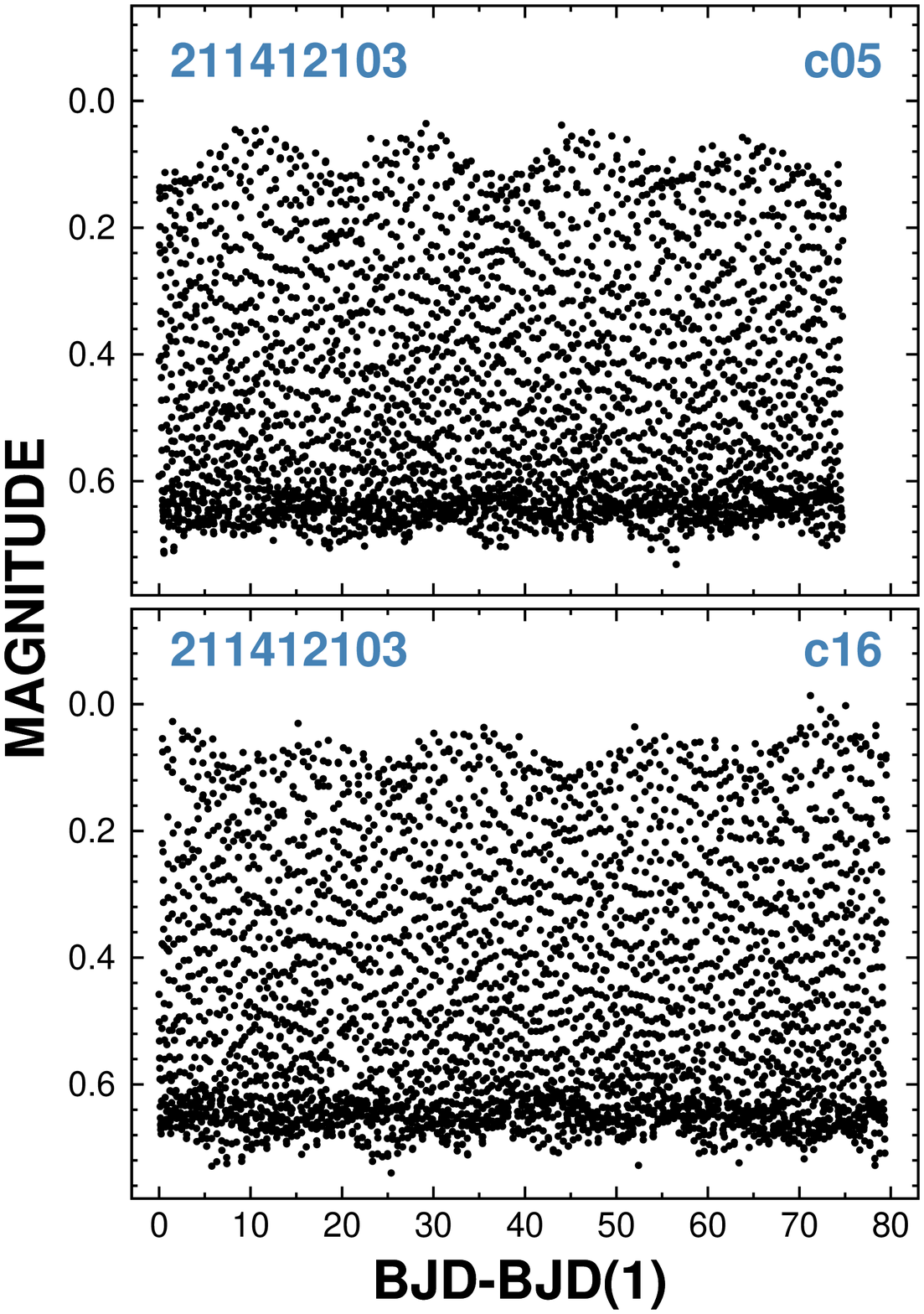}}
    \caption{Residual Fourier spectra of a star from two overlapping 
             campaigns. Spectra of the residuals after subtracting the 
	     variation corresponding to the monoperiodic pulsation are 
	     shown. Insets are zoomed near the fundamental frequency 
	     $\nu_0$. Interestingly, the modulation power is higher 
	     at the harmonics of $\nu_0$. The panels on the right show 
	     the systematics-filtered light curves (the magnitude zero 
	     point is arbitrary). The object was classified as a non-Blazhko 
	     variable by \cite{2019ApJS..244...32P} -- see text 
	     for possible reasons.}
    \label{plachy1}
\end{figure}

In a recent paper, \cite{2019ApJS..244...32P} claim a more 
traditional observed rate of $\sim 50$~\%, based on the analysis 
of the fields of C03$-$C06. Although we did not make a star-by-star 
comparison between their and our results, the following examples 
are aimed for the illustration of the possible sources of the 
discrepancy. 

\begin{figure}[!h]
    \centering
    \subfloat{\includegraphics[angle=-90, width=0.48\textwidth]{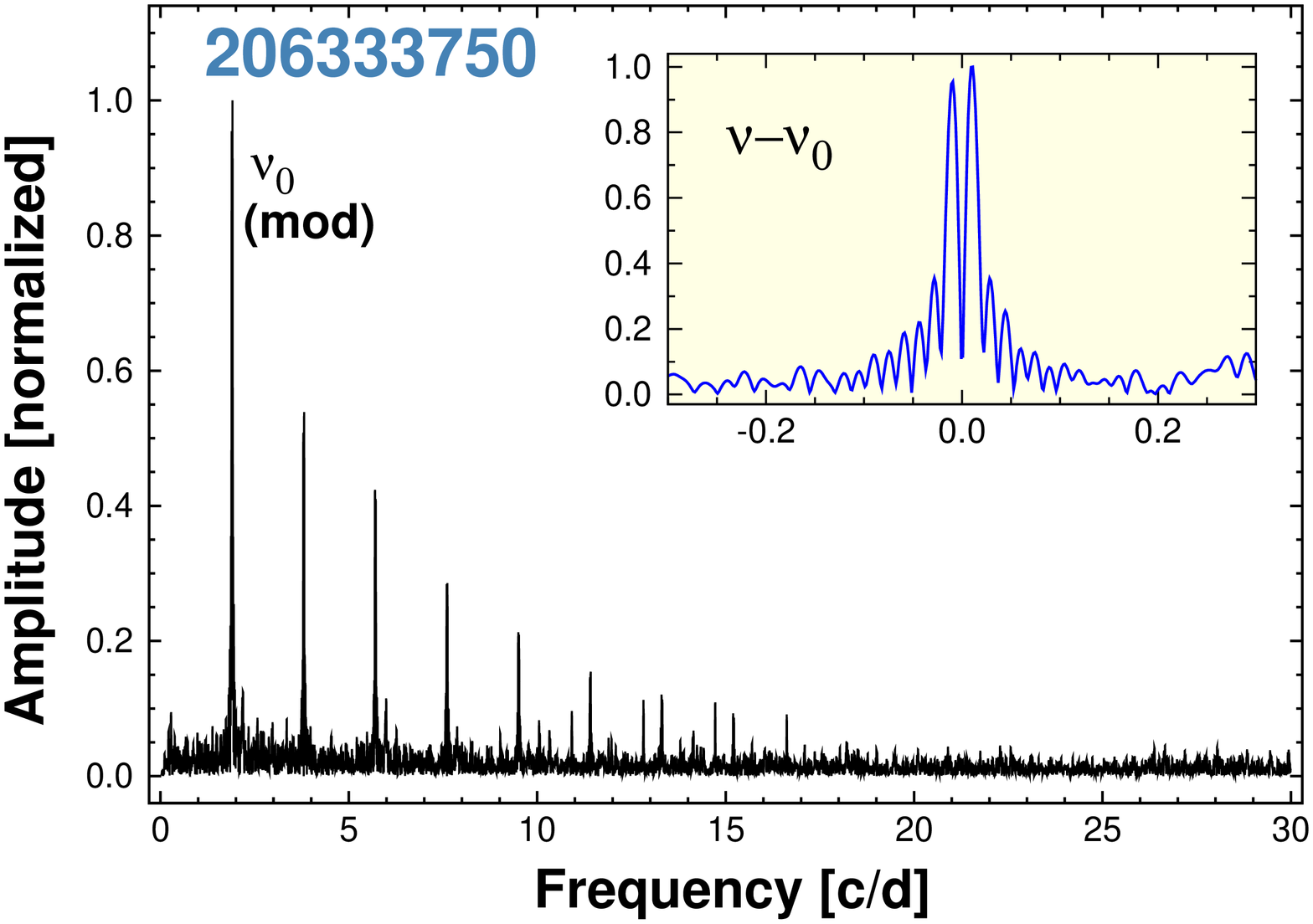}}
    \quad
    \subfloat{\includegraphics[angle=-90, width=0.48\textwidth]{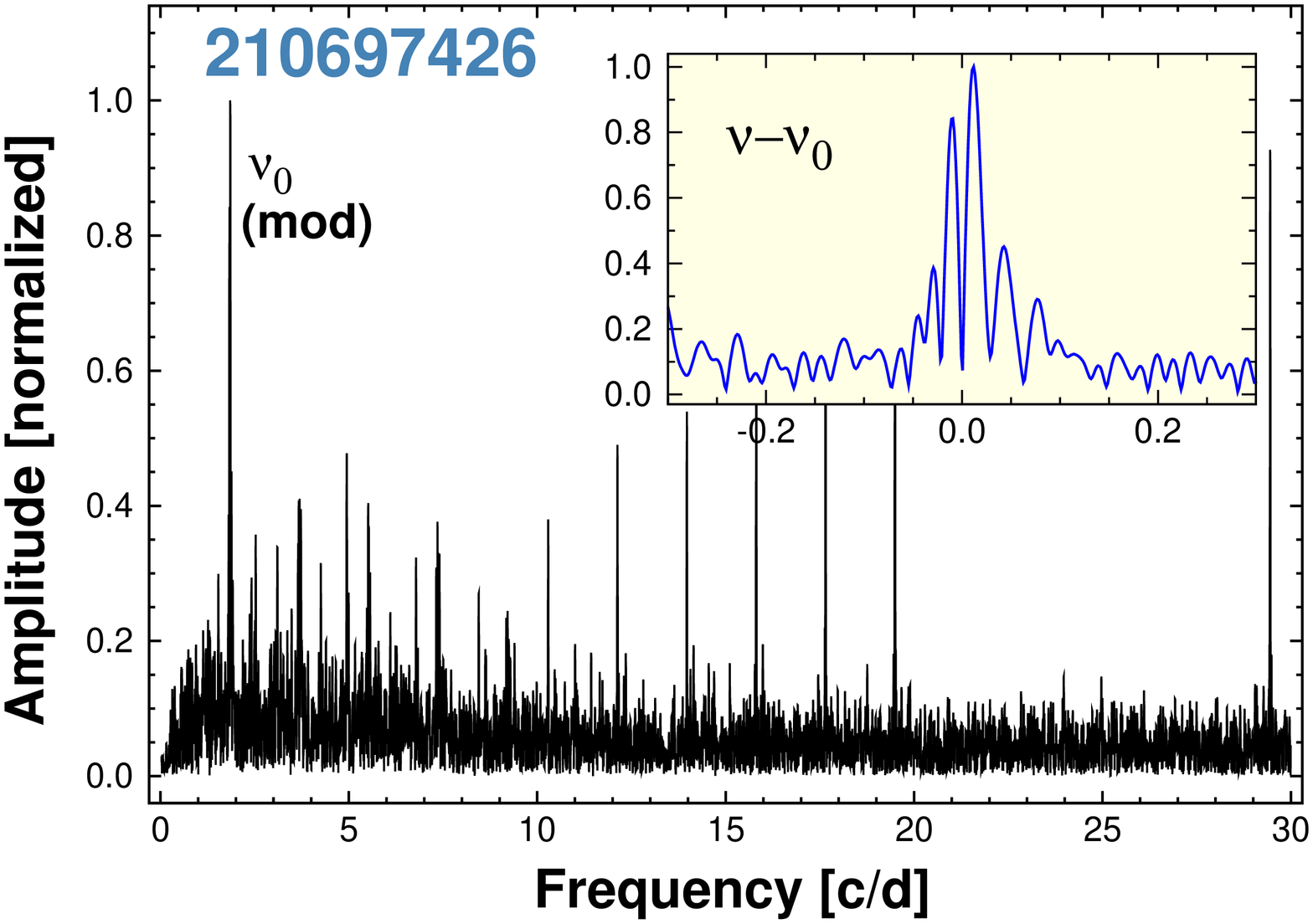}}
    \caption{Fourier spectra of two stars classified as non-Blazhko variables 
             by \cite{2019ApJS..244...32P}. Figure setting is the same 
	     as in Fig.~\ref{plachy1}. High-frequency peaks above $8$~c/d 
	     are aliases (via the half hour sampling) of the higher (16th 
	     and up) harmonics left in the time series. The heights of the 
	     side lobes are $\sim 5$ and $\sim 0.5$~ppt for 206333750 and 
	     210697426, respectively. }
    \label{plachy2}
\end{figure}

Figure~\ref{plachy1} shows an example for a target of short modulation 
period, that was mis-classified likely because of the poor performance 
of the K2SC detrending algorithm in this particular case. Figure~\ref{plachy2},  
on the other hand, shows the frequency spectra of two stars from their 
Fig.~12.\footnote{The 3rd star shown in their figure is not accessible 
from the public archives used in our work.} We guess that in these cases 
K2SC might have overfitted the data and eliminated the small amplitude 
of the long period modulation. We also note that their classification 
(Plachy, private communication) have multiple components (including 
the less sensitive O-C analysis), that leads to the de-selection 
of many candidates that otherwise show signatures of modulation 
in the frequency spectra. 

%
%

%
%
\acknowledgements 
We would like to thank the organizers (and, in particular the Chair of 
the LOC, Karen Kinemuchi) for their effort in making this workshop 
scientifically fruitful and socially commemorable. This research has 
made use of the NASA Exoplanet Archive, which is operated by the 
California Institute of Technology, under contract with the National 
Aeronautics and Space Administration under the Exoplanet Exploration 
Program. Supports from the National Research, Development and Innovation 
Office (grants K~129249 and NN~129075) are acknowledged. 
\end{document}